\documentclass{article}
\input BoxedEPS
\SetRokickiEPSFSpecial  
\HideDisplacementBoxes
\def\hhhh{ \hat{\bar{\psi}} \hat{\bar{\psi}} \hat{\psi} \hat{\psi}}
\def\pppp{ \bar{\psi} \bar{\psi} \psi \psi}
\def\bp{ \bar{\psi} \psi}
\def\sq{\mbox{\scriptsize$\sqcap$\llap{$\sqcup$}}}

\def\DD{{\cal D}}
\def\RT{\mathop{\rm Re \, Tr}}
\def\Tr{\mathop{\rm Tr}}
\def\notp{/ \llap{$p$}}
\def\notA{/ \llap{$A$}}
\def\notpt{/ \llap{$\partial$}}

\begin{document}
\title{Insight into Hadron Structure\\ from Lattice QCD}

\author{J.W. Negele\\
Center for Theoretical Physics\thanks{This work is
supported in part by funds provided by the US Department of Energy (DOE) under
cooperative research agreement \#DF-FC02-94ER40818.}\\ 
Laboratory for Nuclear Science
and Department of Physics\\ 
Massachusetts Institute of Technology, Cambridge
MA 02139\\
MIT-CTP\#2676\quad hep-lat/9709129}

\maketitle

\begin{abstract}
A variety of evidence from lattice QCD is presented revealing the
dominant role of instantons in the propagation of light quarks in the
QCD vacuum and in light hadron structure.  The instanton content of
lattice gluon configurations is extracted, and observables calculated
from the instantons alone are shown to agree well with those
calculated using all gluons.  The lowest 128 eigenfunctions of the
Dirac operator are calculated and shown to exhibit zero modes
localized at the instantons.  Finally, the zero mode contributions to
the quark propagator alone are shown to account for essentially the
full strength of the rho and pion resonances in the vector and
pseudoscalar correlation functions.
\end{abstract}

\section*{Introduction}\noindent
For the nearly three decades since the experimental discovery of quarks and the
formulation of QCD, understanding the essential physics of light quarks in QCD and
the structure of light hadrons has remained an elusive goal.  Since analytic
theoretical techniques are as yet inadequate to solve QCD, a number of
very different QCD-inspired models have been developed that present
quite disparate physical pictures.  For example, non-relativistic quark
models focus on constituent quarks interacting via an adiabatic potential. 
Bag models postulate a region in which relativistic current quarks are
confined and interact by gluon exchange.  Motivated by large $N_c$
arguments, Skyrme models describe the nucleon as a topological soliton
built out of $q \bar{q}$ pairs. Finally instanton models emphasize the role
of topological structures in the vacuum corresponding in the semiclassical
limit to instantons and of the quark zero modes associated with these
topological excitations.

How can one understand which, if any, of these fundamentally different pictures
describes the essential physics of light hadrons? Phenomenology has proven
inconclusive, with each of the models being rich enough that with sufficient
embellishment it can be made to fit the data.  Whereas perturbative QCD has proven
extremely useful in extracting quark and gluon structure functions from
high energy scattering experiments, it is inadequate to understand their
origin.  Hence, it is necessary to turn to nonperturbative methods, and the
only known techniques to solve, rather than model QCD, is lattice field
theory. Our strategy, then, will be to use the fact that lattice calculations
numerically evaluate the path integral for QCD as a tool to identify the
paths that dominate the path integral and thereby identify the physics
that dominates hadron structure.

The lattice results described below indicate that gluons play an extremely important
dynamical role in light hadrons.  Thus, QCD with light quarks is unique among the
many-body systems with which we are familiar in the sense that the quanta
generating the interactions cannot be subsumed into a potential but rather
participate as essential dynamical degrees of freedom. In atoms, for
example, photons play a negligible dynamical role, and to an excellent
approximation may be subsumed into the static Coulomb potential.  In
nuclei, mesons play a minor dynamical role, and to a good approximation
nuclear structure maybe described in terms of two- and three-body
nucleon forces. Indeed, experimentalists need to work very hard and pick
their cases carefully to observe any effects of meson exchange currents.
And in heavy quark systems, much of the physics of $c\bar{c}$ and
$b\bar{b}$ bound states may be understood by subsuming the gluons
into an adiabatic potential with Coulombic and confining behavior. It
turns out that nucleons, however, are completely different in that gluons
are crucial dynamical degrees of freedom.  This result is not entirely
unexpected, since from perturbative QCD, we already know by the work
of Gross and Wilczek\cite{R:JN:01} and Hoodbhoy, Ji and Tang\cite{R:JN:02}
that approximately half (${16}/{3n_f}$ to be precise, where $n_f$ is
the number of active flavors and equals 5 below the top quark mass) of
the momentum and angular momentum comes from glue in the limit of
high $Q^2$.  Furthermore, experiment tells us that this behavior
continues down to non-perturbative scales of the order of several
GeV$^2$.

The physical picture that arises from this work corresponds closely to the physical
arguments and instanton models of Shuryak and
others\cite{R:JN:03,R:JN:04,R:JN:05} in which the zero modes associated with
instantons produce localized quark states, and quark propagation
proceeds primarily by hopping between these states.  The support that
lattice calculations provide for this picture includes quantitative
determination of the instanton content of the QCD vacuum, a comparison
of the effects of all gluon contributions versus those of instantons alone,
direct calculation of the quark zero modes, and demonstration that
these modes dominate the rho and pion contributions 
 to vector and
pseudoscalar correlation functions.

\section*{Background}
\subsection*{Lattice QCD}\noindent
A QCD observable is evaluated by defining quark and gluon
variables on the sites and links of a space-time lattice, writing a Euclidean path
integral of the generic form\cite{R:JN:06} 
\begin{eqnarray}
 \langle Te^{-B\hat{H}} \hhhh \rangle 
&=&Z^{-1} \int  \DD(U) \DD (\bp) e^{-\bar{\psi} M(U) \psi - S(U)} \pppp
\label{E:JN:1} \\ &=& Z^{-1} \int  \DD(U) e^{\ln \det M(U) - S(U)} M^{-1}
(U) M^{-1} (U)
\nonumber
\end{eqnarray}
and evaluating the final integral over gluon link variables $U$ using the
Monte Carlo method.  The link variable is $U=e^{iagA_\mu(x)}$, the Wilson
gluon action is $S(U)=
\frac{2n}{g^2} \sum_{\sq} (1-\frac{1}{N} \RT U_{\sq})$ where
$U_{\sq}$ denotes the product of link variables around a single
plaquette, and $M (U)$ denotes the discrete Wilson approximation to the
inverse propagator $M (U) \to m+ \notpt  + \ ig \, \notA$.   Evolution in
Euclidean time is required to assure a dominantly positive integral and
thus to obtain a statistically accurate Monte Carlo result, and we will
utilize the fact that the Euclidean evolution operator projects out the
lowest energy state having a specified set of quantum numbers:
\begin{equation}
e^{-BH} \psi = \sum_n e^{-BE_n} C_n \psi_n \,\,\,
\mathop{\hbox to 4em{\rightarrowfill}}\limits_{B \gg(E_1-E_0)^{-1}}
\,\,\, e^{-BE_0} C_0 \psi_0\  .
\label{E:JN:2}
\end{equation}

A physical way of understanding the final integral in (\ref{E:JN:1}) is to expand
$M
\equiv (1 + \kappa u)^{-1}$ in powers of the so-called hopping parameter
$\kappa$ which couples neighboring sites with gauge fields, and thereby
generate all the quark time-histories. In the case of propagation of a meson from
$x$ to $y$, the integral for\linebreak[4]
$\langle Te^{-BH} \bar{\psi}(y) \psi(y) \bar{\psi}(x) \psi(x) \rangle$ has the following
three contributions.  Expansion of the two $M^{-1}(U)$ terms generates all
valence quark and antiquark trajectories that begin at the source $x$
and terminate at the sink $y$.  Expansion of  $\ln {\rm Det} M (U)$
generates all disconnected quark loops corresponding to excitation of
quark-antiquark pairs from the Dirac Sea.  Omission of the determinant,
which is very expensive computationally, yields the so-called quenched
approximation in which the quark-antiquark pairs excited from the sea
are neglected.  Finally, when the sum over all plaquettes in $S (U)$ is
expanded out of the exponent, the lattice is tiled in all possible ways by
any number of plaquettes.  After integration over $\int D(U)$, only those
combinations of link variables from expanding $M(U)$ and $S (U)$
survive that correspond to color singlets.  The simplest non-vanishing
tiling for a meson corresponds to completely filling in the region between
the valence quark and antiquark with gluon plaquettes.  If one imagines
cutting this and more and more complicated tilings on a single time slice,
one obtains the physical picture of a quark and antiquark (from cutting
the two valence quark lines generated by $M^{-1}$) connected by a gluon
flux tube (from cutting all the gluon surfaces that connect the quark and
antiquark).  Typical lattices range from $16^3 \times 32$ to $32^3 \times
64$ and thus involve numerical integrations over  $\sim  10^7$
to $10^8$ real variables.

\subsection*{Correlation functions}\noindent
As in the case of other strongly interacting many-body systems, to
understand the structure of the vacuum and light hadrons in
nonperturbative QCD, it is instructive to study appropriately selected
ground state correlation functions, to calculate their properties
quantitatively, and to understand their behavior physically.

The vacuum correlation functions we consider are the point-to-point equal time
correlation functions of hadronic currents
\begin{equation}
R(x) = \langle \Omega |T J(x) \bar{J} (0)|\Omega \rangle
\label{E:JN:3}
\end{equation}
discussed in detail by Shuryak\cite{R:JN:03} and recently calculated in
quenched lattice QCD\cite{R:JN:07}. The motivation for supplementing
knowledge of hadron bound state properties by these correlation
functions is clear if one considers the deuteron. Simply knowing the
binding energy, rms radius, quadruple moment and other ground state
properties yields very little information about the nucleon-nucleon
interaction in each spin, isospin and angular momentum channel as a
function of spatial separation.  To understand the nuclear interaction in
detail, one inevitably would be led to study nucleon-nucleon scattering
phase shifts.  Although, regrettably, our experimental colleagues have
been most inept in providing us with quark-antiquark phase shifts, the
same physical information is contained in the vacuum hadron current
correlation functions $R (x)$.  As shown by Shuryak\cite{R:JN:03}, in many
channels these correlators may be determined or significantly
constrained from experimental data using dispersion relations.  Since
numerical calculations on the lattice agree with empirical results where
available, we regard the lattice results as valid solutions of QCD in all
channels and thus use them to obtain information comparable to
scattering phase shifts.

\begin{figure}[b!] 
\begin{center}
\BoxedEPSF{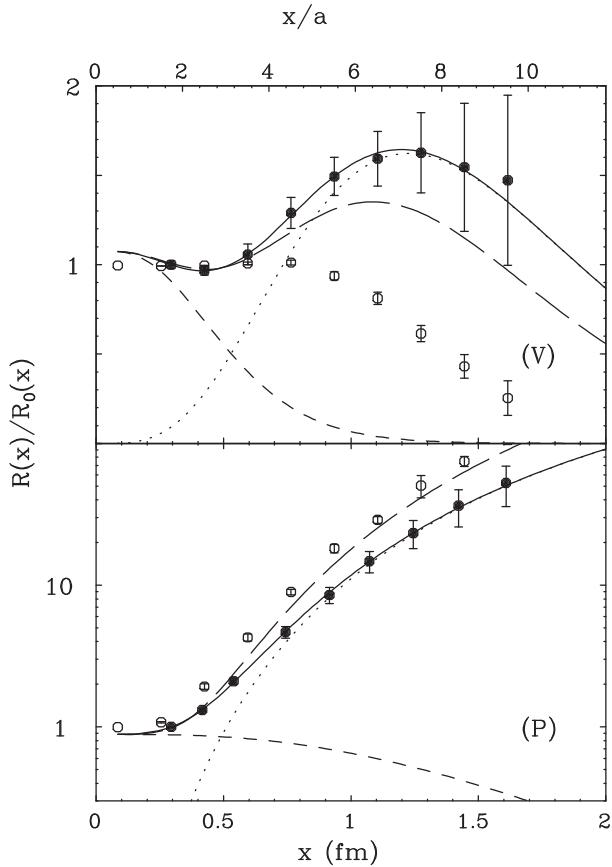}
\medskip
\caption{Vector $(V)$ and 
Pseudoscalar $(P)$ correlation functions are shown in the upper and lower panels
respectively.  Lattice results\protect\cite{R:JN:07} are denoted by the solid points with
error bars and fit by the solid curves, which may be decomposed into continuum and
resonance components denoted by short dashed and dotted curves respectively. 
Phenomenological results determined by dispersion analysis of
experimental data in Ref.~\protect\cite{R:JN:03} are shown by long dashed
curves, and the open circles denote the results of the random instanton
model of Ref.~\protect\cite{R:JN:04}. }
\label{F:JN:1}  
\end{center}
\end{figure}

The correlation functions we calculate in the pseudoscalar, vector,
nucleon and Delta channels are
\begin{eqnarray*}
R(x) &=& \langle \Omega |T J^p(x) \bar{J}^p (0)|\Omega \rangle  \ , \\
R(x) &=& \langle \Omega |T J_\mu(x) \bar{J}_\mu (0)|\Omega \rangle  \ ,
\\ R(x) &=& {\textstyle\frac14} \Tr \left(\langle \Omega |T J^N(x) \bar{J}^N
(0)|\Omega \rangle x_\nu \gamma_\nu \right) \ , \\
\noalign{\hbox{and}}
R(x) &=& {\textstyle\frac14} \Tr \left(\langle \Omega |T J^\Delta_\mu (x)
\bar{J}^\Delta_\mu (0)|\Omega \rangle x_\nu \gamma_\nu  \right) \ , 
\end{eqnarray*}
where
\begin{eqnarray*}
J^p &=& \bar{u} \gamma_5 d \ , \\
J_\mu &=& \bar{u} \gamma_\mu \gamma_5 d \ ,  \\
J^N &=& \epsilon_{abc} [u^a C \gamma_\mu u^b] \gamma_\mu \gamma_5
d^c \ , 
\\
\noalign{\hbox{and}}
J^\Delta_\mu &=& \epsilon_{abc} [u^a C \gamma_\mu u^b] u^c \  . 
\end{eqnarray*}
As in Refs.~\cite{R:JN:03} and  \cite{R:JN:07}, we consider the ratio of the
correlation function in QCD to the correlation function for non-interacting
massless quarks,
${R(x)}/{R_0 (x)}$, which approaches one as $x \to 0$ and displays a
broad range of non-perturbative effects for $x$ of the order of 1~fm. 
Typical results of lattice calculations of ratios of vacuum correlation
functions are shown in Fig.~\ref{F:JN:1}.

Note that the lattice results (solid line) agree well with phenomenological results
from dispersion analysis of data (long dashed curves).  Also, observe that the
vector and pseudoscalar correlation functions are strongly dominated by the rho
and pion contributions (dotted lines) in the region of 0.5 to 1.5 fm.  We will
subsequently show that these rho and pion contributions in turn arise from the
zero mode contributions associated with instantons.

As discussed in Refs.~\cite{R:JN:03,R:JN:07}, these vacuum correlators show strong
indications of instanton dominated physics.  As shown by 't Hooft\cite{R:JN:08},
the instanton induced interaction couples quarks and antiquarks of opposite
chirality leading to strong attractive and repulsive forces in the pseudoscalar and
scalar channels respectively and no interaction to leading order in the vector
channel.  Just this qualitative behavior is observed at short distance in all the
channels we computed.  Furthermore, as shown by the open circles with error
bars   in Fig.~\ref{F:JN:1}, the random instanton model of Shuryak et
al.\null\cite{R:JN:04} reproduces the main features of the correlation functions at
large distance as well.

\subsection*{Instantons}\noindent
The QCD vacuum is understood as a superposition of an infinite number of states of
different winding number, where the winding number characterizes the number of
times the group manifold is covered when one covers the physical space. Just as
there is a stationary point in the action of the Euclidean Feynman path integral
for a double well potential corresponding to the tunneling between the two
degenerate minima, so also there is a classical solution to the QCD equations in
Euclidean time, known as an instanton\cite{R:JN:09}, which describes tunneling
between two vacuum states of differing winding number.  The action associated
with an instanton is
\begin{equation}
S_0 = \frac14 \int d^4 x\, F^a_{\mu \nu} F^a_{\mu \nu} = \frac{48}{g^2
\rho^4} \int d^4 x\, \Bigl(\frac{\rho^2}{x^2 +\rho^2} \Bigr)^4 = \frac{8 \pi^2}{g^2}\  .
\label{E:JN:4}
\end{equation}
Note that the action density has a universal shape characterized by a size $\rho$, and
that the action is independent of $\rho$.  Furthermore, the instanton field strength is
self-dual, {\it i.e.\/} $\tilde{F}^a_{\mu \nu} \equiv \epsilon_{\mu \nu \alpha \beta}
F^a_{\alpha \beta} = \pm F^a_{\mu \nu}$, so that the topological change of an
instanton is
$$
Q \equiv \frac{g^2}{8 \pi^2} {\textstyle\frac14} \int d^4x \tilde{F}^a_{\mu \nu} F^a_{\mu \nu}
= \pm 1\  .
$$
Two features of instantons are particularly relevant to light hadron physics. The
first is the fact that although the fermion spectrum is identical at each minimum
of the vacuum, quarks of opposite chirality are raised or lowered one level
between adjacent minima.   Thus, an instanton absorbs a left-handed quark of
each flavor and emits a right-handed quark of each flavor, and an anti-instanton
absorbs right-handed quarks and emits left-handed quarks.  Omitting heavier
quarks for simplicity, the resulting 't~Hooft interaction involving the operator
$\bar{u}_R u_L
\bar{d}_R d_L \bar{s}_R s_L$ is the natural mechanism to describe otherwise
puzzling aspects of light hadrons.  It is the natural mechanism to flip the helicity of
a valence quark and transmit this helicity to the glue and quark-antiquark pairs,
thereby explaining the so-called ``spin crisis."  It also explains why the two
valence
$u$ quarks in the proton would induce twice as many  $\bar{d} d$ pairs as the
$\bar{u} u$ pairs induced by the single valence $d$ quark.  The second feature is
that each instanton gives rise to a localized zero mode of the Dirac operator
$D_\mu
\gamma_\mu \phi_0 (x) = 0$.  Hence, considering a spectral  representation of the
quark propagator, it is natural that the propagator for the light quarks is
dominated by these zero modes at low energy.  This gives rise to a physical picture
in which
$\bar{q} q$ pairs propagate by ``hopping"  between localized modes
associated with instantons.

\section*{Instanton Content of\\ Lattice Gluon Configurations}
\subsection*{Identifying instantons by cooling}\noindent
The Feynman path integral for a quantum mechanical problem with degenerate
minima is dominated by paths that fluctuate around stationary solutions to the
classical Euclidean action connecting these minima\cite{R:JN:10}. In the case of the
double well potential, a typical Feynman path is composed of segments fluctuating
around the left and right minima joined by segments crossing the barrier.  If one
had such a trajectory as an initial condition, one could find the nearest stationary
solution to the classical action numerically by using an iterative local relaxation
algorithm.  In this method, which has come to be known as cooling, one
sequentially minimizes the action locally as a function of the coordinate on each
time slice and iteratively approaches a stationary solution.  In the case of the
double well, the trajectory approaches straight lines in the two minima joined by
kinks and anti-kinks crossing the barrier and the structure of the trajectory can be
characterized by the number and positions of the kinks and anti-kinks.

In QCD, the corresponding classical stationary solutions to the Euclidean action for the
gauge field connecting degenerate minima of the vacuum are instantons, and we
apply the analogous cooling technique\cite{R:JN:11} to identify the instantons
corresponding to each gauge field configuration. 

\begin{figure}[b!] 
\begin{center}
\BoxedEPSF{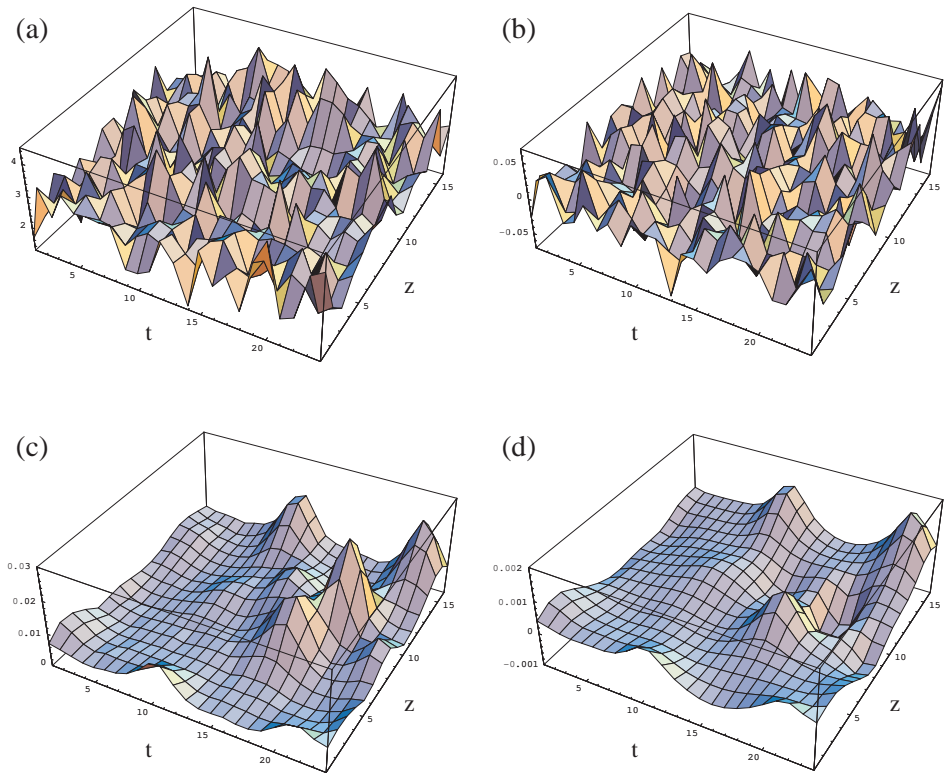}
\medskip
\caption{Instanton content of a typical slice of a gluon configuration at fixed $x$
and $y$ as a function of $z$ and $t$.  The left column shows the action density
$S(1,1,z,t)$ before cooling (a) and after cooling for 25 steps (c).  The right column
shows the topological charge density $Q (1,1,z,t)$ before cooling (b) and after cooling
for 25 steps.}
\label{F:JN:2}  
\end{center}
\end{figure}

The results of using 25 cooling steps as a filter to extract the instanton content of a
typical gluon configuration are shown in Fig.~\ref{F:JN:2}, taken from
Ref.~\cite{R:JN:12} using the Wilson action on a $16^3 \times 24$ lattice at
${6}/{g^2} = 5.7$.  As one can see, there is no recognizable structure before
cooling.  Large, short wavelength fluctuations of the order of the lattice spacing
dominate both the action and topological charge density.  After 25 cooling steps,
three instantons and two anti-instantons can be identified clearly.  The action
density peaks are completely correlated in position and shape with the topological
charge density peaks for instantons and with the topological charge density valleys
for anti-instantons.  Note that both the action and topological charge densities are
reduced by more than two orders of magnitude  so that the fluctuations removed by
cooling are several orders of magnitude larger than the topological excitations that
are retained.

Setting the coupling constant, or equivalently, the lattice spacing, and quark mass by
the nucleon and pion masses in the usual way, it turns out that the characteristic size
of the instantons identified by cooling is 0.36~fm and the density is 1.6 fm$^{-4}$, in
reasonable agreement with the value of 0.33~fm and 1.0~fm$^{-4}$ in the liquid
instanton model\cite{R:JN:04}.

\subsection*{Comparison of results with all gluons\\ and with only instantons}\noindent
One dramatic indication of the role of instantons in light hadrons is to
compare observables calculated using all gluon contributions with those obtained
using only   the instantons remaining after cooling.  Note that there are truly dramatic
differences in the gluon content before and after cooling.  Not only has the action
density decreased by two orders of magnitude, but also the string tension has
decreased to 27\% of its original value and the Coulombic and magnetic hyperfine
components of the quark-quark potential are essentially zero.  Hence, for example,
the energies and wave functions of charmed and $B$ mesons would be drastically
changed.

\begin{figure}[b!] 
\begin{center}
\BoxedEPSF{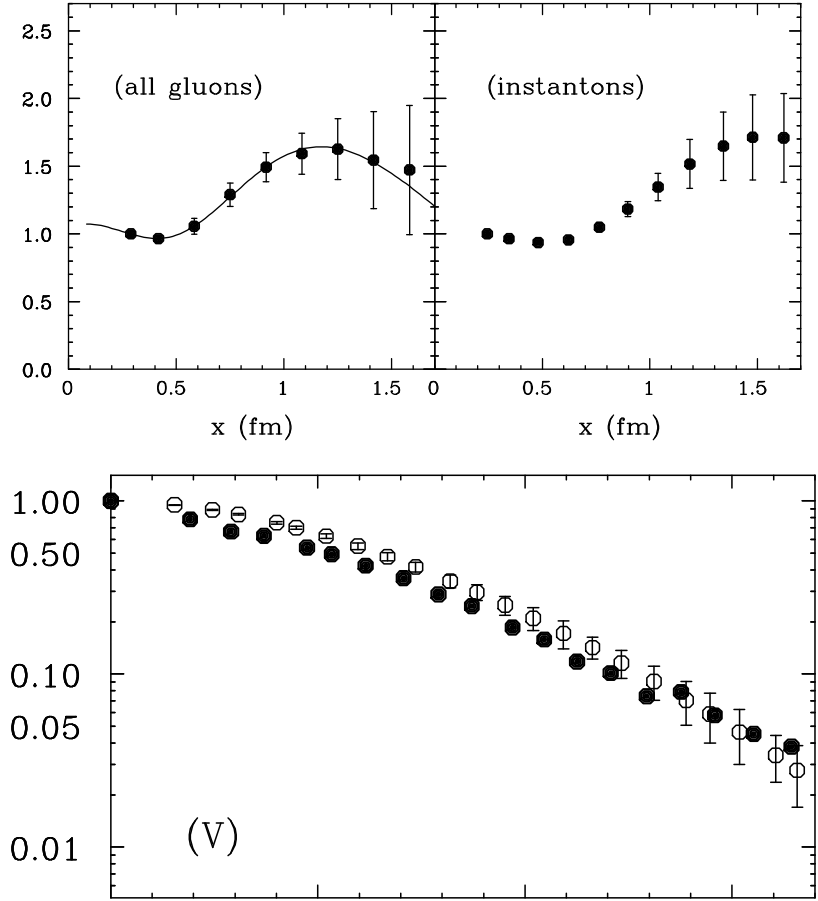}
\caption{Comparison of rho observables calculated with all gluon
configurations and only instantons. The upper left-hand plot shows the vacuum
correlator in the rho channel calculated with all gluons as in
Fig.~\protect\ref{F:JN:1} and the upper right-hand plot shows the analogous result
with only instantons.  The lower plot shows the ground state density-density
correlation function for the rho with all gluons (solid circles) and with only
instantons (open circles).  Error bars for the solid circles are comparable to the
open circles and have been suppressed for clarity.}
\label{F:JN:3}  
\end{center}
\end{figure}

As shown in Fig.~\ref{F:JN:3}, however, the properties of the rho meson are virtually
unchanged.  The vacuum correlation function in the rho (vector) channel and the
spatial distribution of the quarks in the rho ground state, given by the ground state
density-density correlation function\cite{R:JN:13} $\langle \rho| \bar{q} \gamma_0
q(x) \bar{q} \gamma_0 q(0) | \rho \rangle$, are statistically indistinguishable before
and after cooling.  Also, as shown in Ref.~\cite{R:JN:11}, the rho mass is unchanged
within its 10\% statistical error.  In addition, the pseudoscalar, nucleon, and
delta vacuum correlation functions and nucleon and pion density-density correlation
functions are also qualitatively unchanged after cooling, except for the removal of the
small Coulomb induced cusp at the origin of the pion.

Although these cooling studies strongly indicate that instantons play an essential role
in light quark physics, cooling has the disadvantage of modifying the instanton
content of the original gluon configuration. It is possible to avoid the gradual
shrinkage of a single instanton until it eventually falls through the lattice by using
an improved action that is sufficiently scale independent\cite{R:JN:14}.  However,
pairs of instantons and anti-instantons will eventually attract each other and
annihilate, thereby continually eroding the original distribution.  Hence,  it is
valuable to complement these cooling calculations by studies of the zero modes
associated with instantons, which, as we show in the next section, can be carried out
successfully on the original uncooled gluon configurations.

\section*{Quark Zero Modes and Their\\ Contributions to Light
Hadrons}\nopagebreak
\subsection*{Eigenmodes of the Dirac operator}\noindent
In the continuum limit, the Dirac operator for Wilson fermions approaches the
familiar continuum result 
$$ 
D \psi_x = \psi_x - \kappa \sum_\mu \Bigl[ (r-\gamma_\mu) u_{x,\mu}
\psi_{x+\mu} + (r + \gamma_\mu) u^\dagger_{x-\mu,\mu} \psi_{x-\mu} \Bigr] \to
\frac{1}{m} \Bigl[ m+i (\notp + g \, \notA)\Bigr] \psi\  .
$$

In the free case, the continuum spectrum is $\frac{1}{m}[m+i |\vec{p}|]$ and the
Wilson lattice operator approximates this spectrum in the physical regime and
pushes the unphysical fermion modes to  very large (real) masses.   In the presence
of an instanton of size $\rho$ at $x=0$, it is shown in Ref.~\cite{R:JN:15} that the lattice
operator produces a mode with zero imaginary part that approaches the
continuum result 
$$ 
\psi_0(x)_{s,\alpha} = u_{s, \alpha} \frac{\sqrt{2}}{\pi} \frac{\rho}{(x^2 +
\rho^2)^{3/2}}
$$
and whose mixing with other modes goes to zero as the lattice volume goes to
infinity.  In addition, instanton-anti-instanton pairs that interact sufficiently form
complex conjugate pairs of eigenvalues that move slightly off the real axis.  Thus,
by observing the Dirac spectrum for a lattice gluon configuration containing a
collection of instantons and anti-instantons, it is possible to identify zero modes
directly in the spectrum.

\begin{figure}[b!] 
\begin{center}
\BoxedEPSF{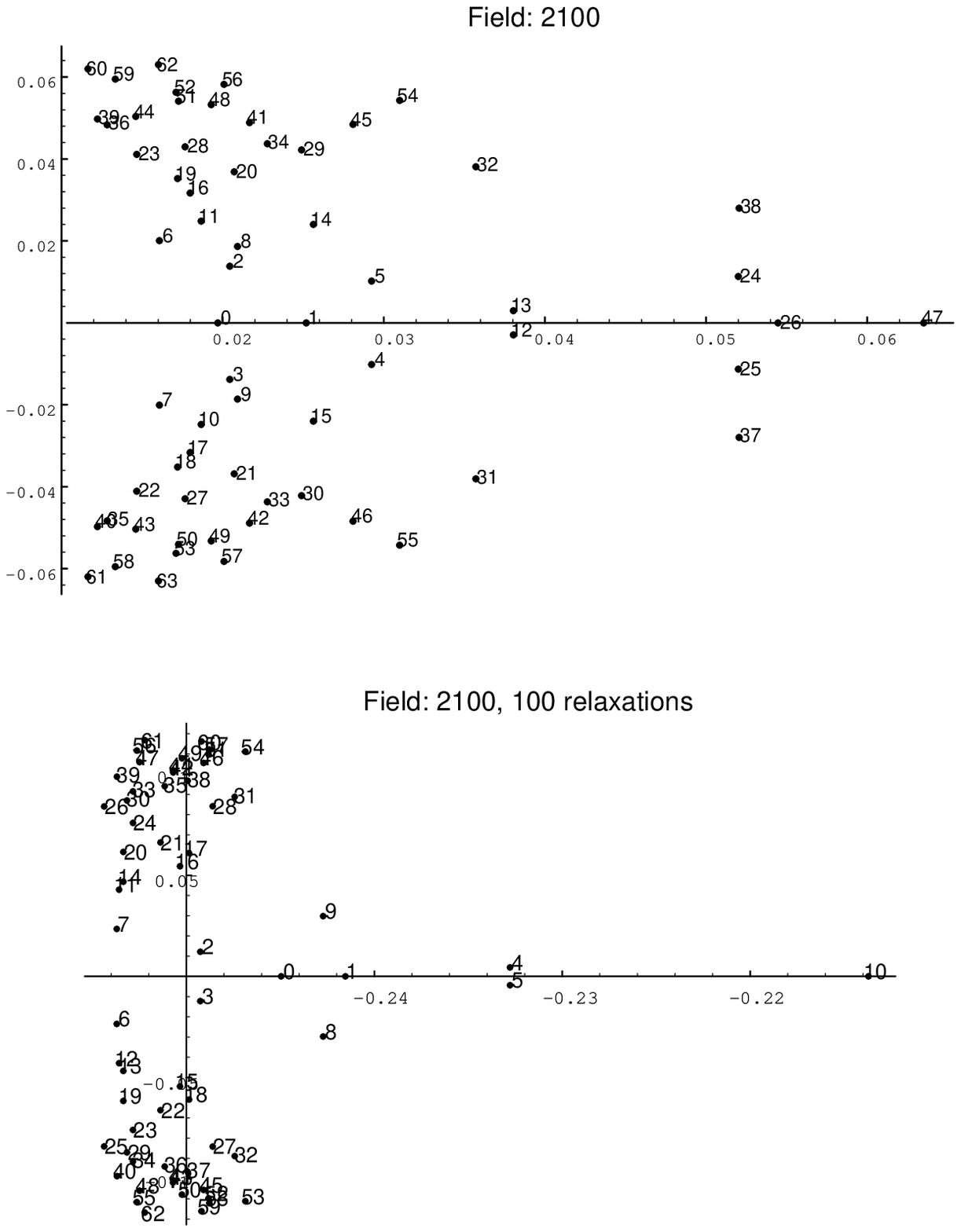}
\medskip
\caption{Lowest 64 complex eigenvalues of the Wilson--Dirac operator for an
unquenched gluon configuration both before (upper plot) and after cooling (lower
plot).  The scale is such that  0.06 on the imaginary
axis roughly corresponds to the lowest Matsubara frequency, 380 MeV.}
\label{F:JN:4}  
\end{center}
\end{figure}

\begin{figure}[b!] 
\begin{center}
\BoxedEPSF{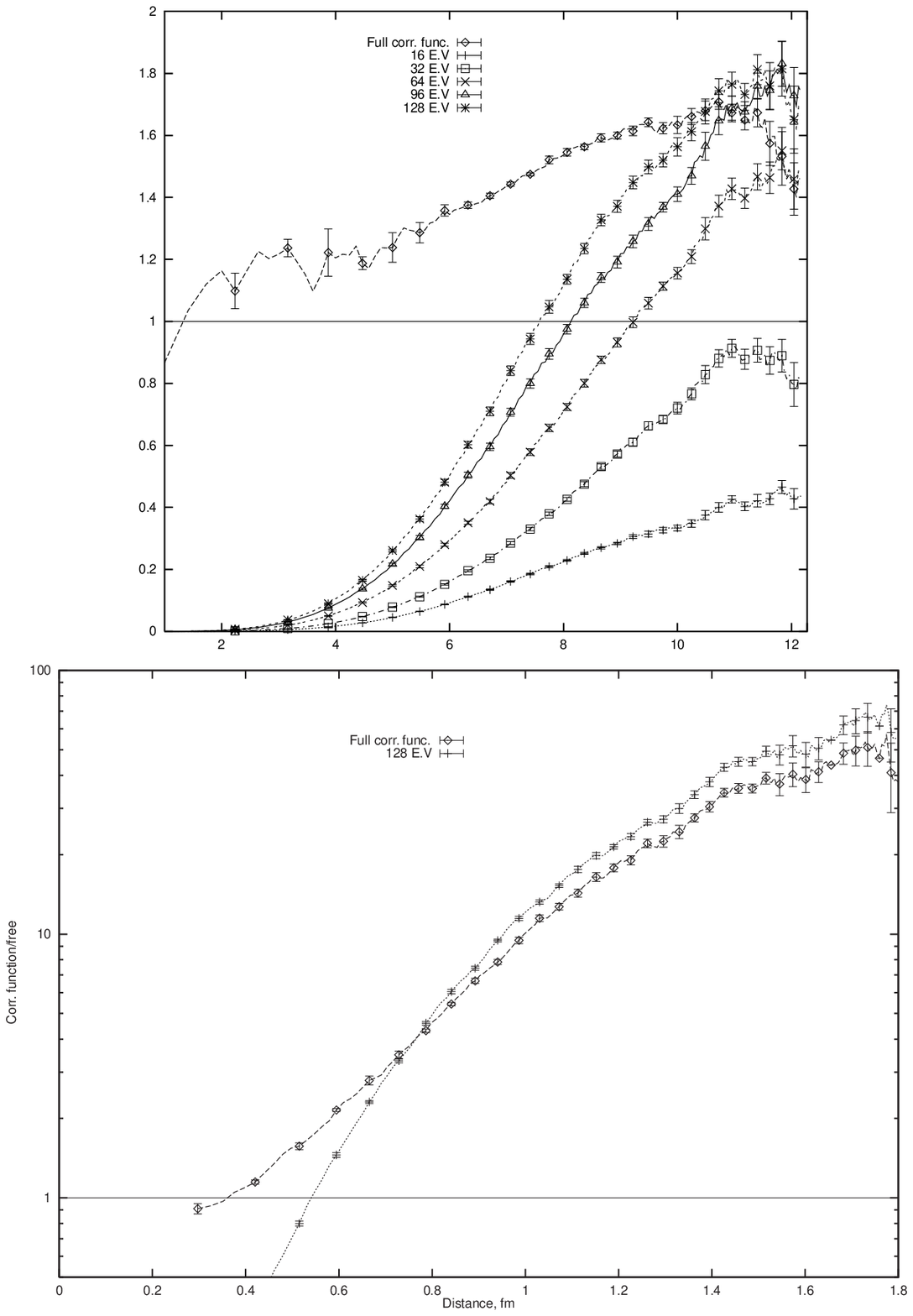 scaled 950}
\caption{Contributions of low Dirac eigenmodes to the vector (upper graph)
and pseudoscalar (lower graph) vacuum correlation functions.  The upper
graph shows the contributions of 16, 32, 64, 96, and 128 eigenmodes compared with
the full correlation function for an unquenched configuration with a 63~MeV valence
quark mass.  The lower graph compares 128 eigenmodes with the full correlation
function for a quenched configuration with a 23~MeV quark mass.}
\label{F:JN:5}  
\end{center}
\end{figure}

Fig.~\ref{F:JN:4} shows the lowest 64 complex eigenvalues of the Dirac operator on
a $16^4$ unquenched gluon configuration for ${6}/{g^2}=5.5$ and $\kappa=0.16$,
both before and after cooling (where 100 relaxation steps with a parallel algorithm
are comparable to 25 cooling steps).  The lower, cooled, plot has just the structure
we expect with a number of isolated instantons with modes on the real axis and
pairs of interacting instantons slightly off the real axis.  However, even though the
uncooled case shown in the upper plot also contains fluctuations several orders of
magnitude larger than the instantons (as seen in Fig.~\ref{F:JN:2}), it shows the
same structure of isolated instantons and interacting pairs.  To set the scale, note
that if we had antiperiodic boundary conditions in time, the lowest Matsubara
mode ($ip = i\frac{\pi}{L}$) would occur at 0.06 on the imaginary axis, so all the
modes below this value are presumably the results of zero modes.

\subsection*{Zero mode expansion}\noindent
The Wilson--Dirac operator has the property that $D=\gamma_5 D^\dagger
\gamma_5$, which implies that $\langle \psi_j |\gamma_5| \psi_i \rangle=0$ unless
$\lambda_i = \lambda^*_j$ and we may write the spectral representation of the
propagator
$$
\langle x |D^{-1}|y \rangle = \sum_i \frac{\langle x|\psi_i \rangle \langle \psi_{\bar{i}}
|\gamma_5| y\rangle}{\langle \psi_{\bar{i}}|\gamma_5| \psi_i\rangle \lambda_i}
$$
where $\lambda_i = \lambda_{\bar{i}}^*$.  A clear indication of the role of zero modes
in light hadron observables is the degree to which truncation of the expansion to the
zero mode zone reproduces the result with the complete propagator.

Fig.~\ref{F:JN:5} shows the result of truncating the vacuum correlation functions
for the vector and pseudoscalar channels to include only low
eigenmodes\cite{R:JN:15}. On a $16^4$ lattice, the full propagator contains
786,432 modes. The top plot of Fig.~\ref{F:JN:5}  shows the result of including the
lowest 16, 32, 64, 96, and finally 128 modes. Note that the first 64 modes
reproduce most of the strength in the rho resonance peak pointed out in
Fig.~\ref{F:JN:1}, and by the time we include the first 128 modes, all the strength
is accounted for.  Similarly, the lower plot in Fig.~\ref{F:JN:5} shows that the
lowest 128 modes also account for  the analogous pion contribution to the
pseudoscalar vacuum correlation function.  Thus, without having to resort to
cooling, by looking directly at the contribution the lowest eigenfunctions, we have
shown that the zero modes associated with instantons dominate the propagation
of rho and pi mesons in the QCD vacuum.

\subsection*{Localization}\noindent
Finally, it is interesting to ask whether the lattice zero mode eigenfunctions are
localized on instantons.  
  This was studied by plotting the quark density distribution for
  individual eigenmodes in
  the $x$-$z$ plane for all values of $y$ and~$t$, and comparing with
  analogous plots of the action density. As expected, for a cooled
  configuration the eigenmodes 
correspond to
linear combinations of localized zero modes at each of the instantons.
(Because there are no symmetries, the coefficients are much more
complicated than the even and odd combinations in a double well or the
Bloch waves in a periodic potential.)  What is truly remarkable, however, is
 that the eigenfunctions of the uncooled configurations also
exhibit localized peaks at locations at which instantons are identified by
cooling.  Thus, in spite of the fluctuations several orders of magnitude larger
than the instanton fields themselves, the light quarks essentially average
out these fluctuations and produce localized peaks at the topological
excitations.  When one analyzes a number of eigenfunctions, one finds that
all the instantons remaining after cooling correspond to localized quark
fermion peaks in some eigenfunctions.  However, some fermion peaks are
present for the initial gluon configurations that do not correspond to
instantons that survive cooling.  These presumably correspond to
instanton--anti-instanton pairs that were annihilated during cooling.

\section*{Conclusion}\noindent
Altogether, the lattice calculations reported here provide strong evidence
that instantons play a dominant role in quark propagation in the vacuum
and in light hadron structure.  We have shown that the instanton content of
gluon configurations can be extracted by cooling, and that the instanton size
and density is consistent with the instanton liquid model.  We obtain
striking agreement between vacuum correlation functions, ground state
density-density correlation functions, and masses calculated with all gluons
and with only instantons.  Zero modes associated with instantons are clearly
evident in the Dirac spectrum, and  account for the rho and pi contributions
to vector and pseudoscalar vacuum correlation functions.  Finally, we have
observed directly quark localization at instantons in uncooled configurations.

\subsubsection*{Acknowledgments}\noindent
It is a pleasure to acknowledge the essential role of Richard Brower, Ming Chu, Jeff
Grandy, Suzhou Huang, Taras Ivananko, Kostas Orginos, and Andrew Pochinsky who
collaborated in various aspects of this work.  We are also grateful for the donation by
Sun Microsystems of the 24 Gflops E5000 SMP cluster on which the most recent
calculations were performed and the computer resources provided by NERSC with
which this work was begun.

\end{document}